\documentstyle[epsfig]{aipproc}

\begin{document}
\thispagestyle{empty}
\newcommand{\ti}{\tilde}
\hfill DESY 97-097

\hfill IFT 4/97


\phantom{X} \vspace{2.3cm}

\begin{center}
{\Large\bf New Physics at HERA: Implications\\[2mm] for $e^+e^-$
                               Scattering at LEP2}
\vskip .5cm
{\large  J. Kalinowski} 
\vskip .3cm
{\it Deutsches Elektronen-Synchrotron DESY, D-22607 Hamburg\\and\\
Institute of Theoretical Physics, Warsaw University, PL-00681 Warsaw}
\end{center}
\vskip 3.cm

\begin{center}
{\bf Abstract}
\end{center}

The impact of virtual leptoquark or $R$-parity breaking squark
exchange as well as generic contact interactions on the production of
quark--antiquark pairs in $e^+e^-$ annihilation, in particular at
LEP2, is summarized. An exciting possibility of sneutrino formation in
$e^+e^-$ scattering is also mentioned.

\vspace{6.5cm}
\noindent
{\it Talk presented at the  5th International Workshop on 
``Deep Inelastic Scattering and QCD'' 
(DIS '97), Chicago, Illinois, USA, April 14-18, 1997.}

\vfill

\setcounter{page}{0} 
\newpage

\title{New Physics at HERA: Implications for $e^+e^-$
                               Scattering at LEP2}

\author{J. Kalinowski}
\address{Deutsches Elektronen-Synchrotron DESY, D-22607 Hamburg\\and\\
Institute of Theoretical Physics, Warsaw University, PL-00681 Warsaw}

\maketitle

\begin{abstract}
  The impact of virtual leptoquark or $R$-parity breaking squark
  exchange as well as generic contact interactions on the production of
  quark--antiquark pairs in $e^+e^-$ annihilation, in particular at
  LEP2, is summarized.  An exciting possibility of sneutrino
  formation in $e^+e^-$ scattering is also mentioned.
\end{abstract}

\def\lsim{\:\raisebox{-0.5ex}{$\stackrel{\textstyle<}{\sim}$}\:}
\def\gsim{\:\raisebox{-0.5ex}{$\stackrel{\textstyle>}{\sim}$}\:}

\newcommand{\newc}{\newcommand}
\newc{\pbi}{pb$^{-1}$}
\newc{\ra}{\rightarrow}
\newc{\ee}{$e^+e^-$\ }
\newc{\qq}{$q\bar{q}$\ }
\newc{\dd}{$d\bar{d}$\ }
\newc{\uu}{$u\bar{u}$\ }
\newc{\lqs}{$LQ/\tilde{q}$\ }
\newc{\eeqq}{$e^+e^-\ra q\bar{q}$\ }
\newc{\eeuu}{$e^+e^-\ra u\bar{u}$\ }
\newc{\eedd}{$e^+e^-\ra d\bar{d}$\ }
\newc{\beq}{\begin{eqnarray}}
\newc{\eeq}{\end{eqnarray}}
\newc{\dqu}{\delta_{qu}}
\newc{\dqd}{\delta_{qd}}
\newc{\non}{\nonumber}
\newc{\noi}{\noindent}
\newcommand{\vs}{\vspace{5mm}}
\newcommand{\ecms}{\sqrt{s}}
%
%
\def\tp{these proceedings}
\def\ib#1,#2,#3{       {\it ibid.\/ }{\bf #1} (19#2) #3}
\def\ap#1,#2,#3{       {\it Ann.~Phys.~(NY)\/ }{\bf #1} (19#2) #3}
\def\ijmp#1,#2,#3{     {\it Int.\ J.~Mod.\ Phys.\/ } {\bf A#1} (19#2) #3}
\def\mpl#1,#2,#3 {     {\it Mod.~Phys.~Lett.\/ } {\bf A#1} (19#2) #3}
\def\npb#1,#2,#3{       {\it Nucl.\ Phys.\/ }{\bf B#1} (19#2) #3}
\def\npps#1,#2,#3{     {\it Nucl.\ Phys.~B (Proc.~Suppl.)\/ }{\bf B#1}
                             (19#2) #3}
\def\plb#1,#2,#3{      {\it Phys.\ Lett.\/ }{\bf B#1} (19#2) #3}
\def\pr#1,#2,#3{       {\it Phys.\ Rev.\/ }{\bf #1} (19#2) #3}
\def\prd#1,#2,#3{      {\it Phys.\ Rev.\/ }{\bf D#1} (19#2) #3}
\def\prep#1,#2,#3{     {\it Phys.\ Rep.\/ }{\bf #1} (19#2) #3}
\def\prl#1,#2,#3{      {\it Phys.\ Rev.\ Lett.\/ }{\bf #1} (19#2) #3}
\def\pro#1,#2,#3{      {\it Prog.~Theor.\ Phys.\/ }{\bf #1} (19#2) #3}
\def\rmp#1,#2,#3{      {\it Rev.~Mod.~Phys.\/ }{\bf #1} (19#2) #3}
\def\sp#1,#2,#3{       {\it Sov.~Phys.~Usp.\/ }{\bf #1} (19#2) #3}
\def\zpc#1,#2,#3{      {\it Z.~Phys.\/ }{\bf C#1} (19#2) #3}
\def\appb#1,#2,#3{     {\it Acta Phys.\ Polon.\/ }{\bf B#1} (19#2) #3}

Recently both HERA experiments reported an excess of events in
positron--proton scattering at very high $Q^2$ values \cite{sem}.
Unambiguous interpretation of these events is not possible at present
due to limited statistics.  Nevertheless, interesting explanations in
terms of new physics, either as contact terms in the effective
Lagrangian or leptoquarks ($LQ$) and/or squarks with $R$-parity
violating couplings ($\tilde{q}$) with masses $m_{LQ/\tilde{q}} \sim
200$ GeV, have been studied in detail.  If true, a large variety of
phenomena are expected to be observed experimentally in other
reactions. If leptoquarks/squarks exist, they can be pair produced in
$p\bar{p}$, \ee and $\gamma\gamma$ collisions; single production with
leptons or quarks can also be explored in these and other reactions.
Below the production threshold, indirect effects generated by the
exchange of virtual $LQ/\tilde{q}$, or by contact terms, are important
means to explore the nature of these new physics interpretations.  A
general classification of leptoquarks respecting the SM symmetries has
been presented in Ref.\ \cite{BRW}, assuming baryon- and lepton-number
conserving, family diagonal and chiral Yukawa couplings to
lepton--quark pairs to avoid restrictions derived from proton decay
and low-energy experiments.  Only a small subset of these states is
realized in supersymmetric theories with $R$-parity breaking coupled
via a term $\lambda'_{ijk}L^i_LQ^j_L\bar{D}^k_R $ in the
superpotential.

The interpretation of HERA data in terms of contact interactions,
leptoquarks or squarks have been summarized in separate talks
\cite{inn}.  In this talk implications of possible interpretations of
HERA events on \eeqq annihilation \cite{alt} are summarized. This
process is mediated by $\gamma,Z$ exchanges in the $s$-channel, and
$LQ/\tilde{q}$ exchanges in the $t/u$-channels; $LQ/\tilde{q}$ with
the fermion number $F=0$ are exchanged in the $t$-, with $F=2$ in the
$u$-channel.  Both scalar ($S_I$) and vector ($V_I$) leptoquarks of
isospin $I$, and squarks ($\tilde{q}$) are considered.

After the Fierz transformation, the $t/u$-channel $LQ/\tilde{q}$
exchange amplitudes in \eeqq generate only {\it (lepton vector
  current)}$\times${\it (quark vector current)} terms in addition to
the standard $s$-channel amplitudes.  This leads to a convenient
representation of the matrix elements including a transparent
interference pattern of $LQ/\tilde{q}$ with $SM$ $\gamma/Z$ exchanges.
Leptoquarks with $I=0,1$ contribute to equal-helicity $LL$ and $RR$
amplitudes, while leptoquarks with $I=1/2$ contribute to
opposite-helicity amplitudes $RL$ and $LR$, where the first (second)
index denotes the helicity of incoming electron (outgoing quark).  All
$F=0$ leptoquarks/squarks contribute with the same positive sign, all
$F=2$ with the negative sign.  For a given $F$, the sign of the
interference with $\gamma/Z$ exchange is determined by the sign of the
generalized charges $Q_{ik}^{eq}= e^2\,Q_eQ_q
+{g^e_ig^q_k}/(1-m_Z^2/s)$, $i,k=L,R$. In the energy range of LEP2
they are negative for $u$-quarks and positive for $d$-quarks, except
$Q_{RL}^{ed}$ which is negative. [The left/right $Z$ charges of the
fermions are defined as $ g^f_L=e\,(I^f_3-s^2_W Q_f)/s_Wc_W$,
$g^f_R=-e\, s^2_W Q_f/s_Wc_W$ with $s_W=\sin\Theta_w$,
$c_W=\cos\Theta_w$].

\vspace{1.5cm}\hspace*{3mm}
\begin{figure}[htb] 
\unitlength 0.43mm
\begin{picture}(10,100)(0,0)
\put(3,-1){\epsfxsize=6.815cm \epsfysize=6.28cm \epsfbox{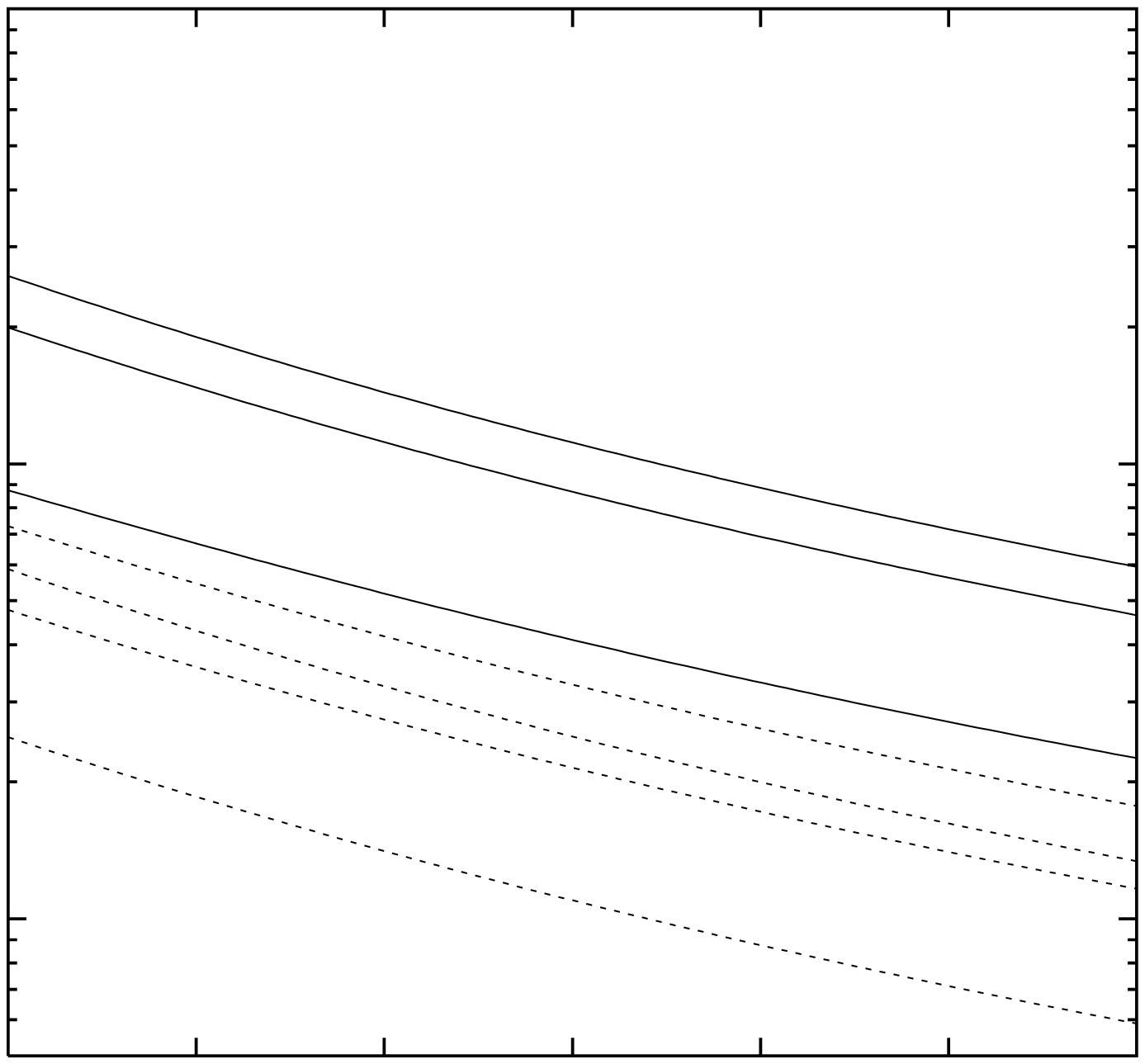}}
\put(30,125){\makebox(0,0)[l]{\small $\Delta = \frac{\sigma(SM\oplus 
LQ)} {\sigma(SM)}-1$}}
\put(100,110){\makebox(0,0)[l]{\small scalar $LQ  $ }}
\put(60,72){\makebox(0,0)[l]{\scriptsize $-\tilde{S}_{1/2}^L$}}
\put(35,87){\makebox(0,0)[l]{\scriptsize $+ S_{1/2}^L$}}
\put(50,100){\makebox(0,0)[l]{\scriptsize $+ S_{1/2}^R$}}
\put(33,37){\makebox(0,0)[l]{\scriptsize $+\tilde{S}_0^R$}}
\put(65,43){\makebox(0,0)[l]{\scriptsize $-S_0^R$}}
\put(90,48){\makebox(0,0)[l]{\scriptsize $+S_1^L$}}
\put(108,50){\makebox(0,0)[l]{\scriptsize $-S_0^L$}}
\put(80,20){\makebox(0,0){\small $m_{LQ}$ [GeV]}}
\put(153,5){\makebox(0,0){\small 500}}
\put(108,5){\makebox(0,0){\small 400}}
\put(63,5){\makebox(0,0){\small 300}}
\put(18,5){\makebox(0,0){\small 200}}
\put(15,30){\makebox(0,0)[r]{\small $10^{-3}$}}
\put(15,85){\makebox(0,0)[r]{\small $10^{-2}$}}
\put(15,140){\makebox(0,0)[r]{\small $10^{-1}$}}
\put(20,20){\makebox(0,0)[l]{(a)}}
\end{picture}
\begin{picture}(100,100)(-150,0)
\put(3,-1){\epsfxsize=6.815cm \epsfysize=6.28cm \epsfbox{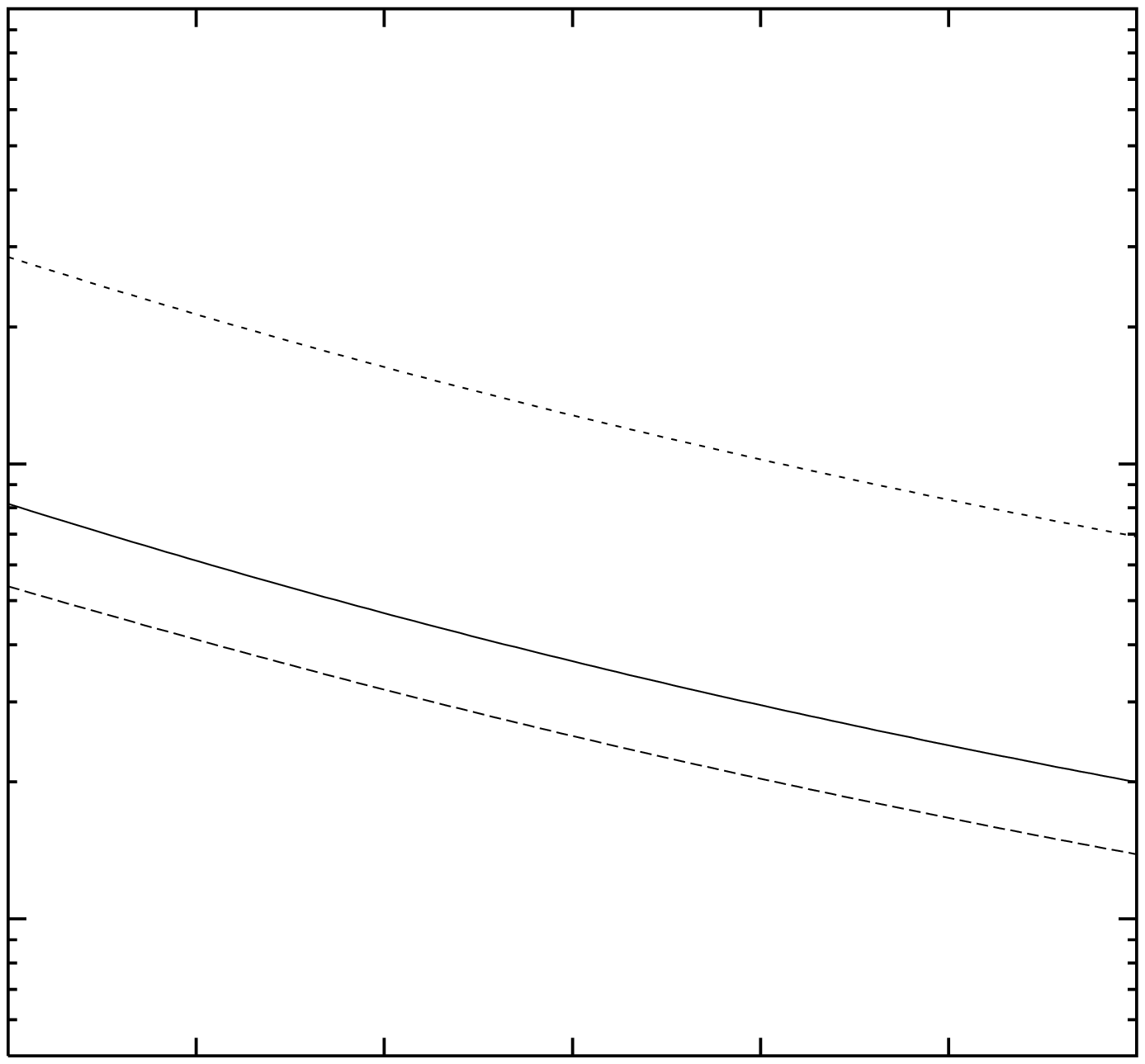}}
\put(30,125){\makebox(0,0)[l]{\small $\Delta = \frac{\sigma(SM\oplus 
\tilde{q})}{\sigma(SM)}-1$}}
\put(100,110){\makebox(0,0)[l]{\small $\lambda'=0.1$}}
\put(40,77){\makebox(0,0)[l]{\footnotesize 
             $-\Delta$: $ \sum q\bar{q}$}}
\put(40,108){\makebox(0,0)[l]{\footnotesize $-\Delta$: $c\bar{c}$}}
\put(40,47){\makebox(0,0)[l]{\footnotesize $-\Delta$: 
$d\bar{d}$, $s\bar{s}$}}
\put(100,20){\makebox(0,0)[l]{\small $m_{\tilde{q}}$ [GeV]}}
\put(152,5){\makebox(0,0){\small 500}}
\put(108,5){\makebox(0,0){\small 400}}
\put(63,5){\makebox(0,0){\small 300}}
\put(18,5){\makebox(0,0){\small 200}}
\put(15,30){\makebox(0,0)[r]{\small $10^{-3}$}}
\put(15,85){\makebox(0,0)[r]{\small $10^{-2}$}}
\put(15,140){\makebox(0,0)[r]{\small $10^{-1}$}}
\put(20,20){\makebox(0,0)[l]{(b)}}
\end{picture}
\caption{ Impact of (a) leptoquark [solid lines scaled by factor 10], 
  (b) squark exchange in \eeqq 
  for $\protect\sqrt{s} = 192$ GeV [adapted from J. Kalinowski et al., 
DESY 97-038, hep-ph/9703288].}
\label{figxsq}
\end{figure}

If the HERA events are interpreted as the signal of \lqs production
with $F=0$ generated in $e^+$--valence-quark collisions, the Yukawa
coupling is of the order 0.05, {\it i.e.} $\sim e/10$. Then the
$t/u$-channel exchange of a leptoquark affects the \eeuu or \eedd
parton cross sections generally only at the level of a percent (up to
10\,\% for $V_0$ and $V_1$).  Summing up all parton channels, the
impact is slightly smaller.  In Fig.1a the sensitivity of the total
hadronic \ee cross section to the entire ensemble of scalar
leptoquarks (superscripts in the figure denote the chirality of the
Yukawa coupling) is shown for the Yukawa coupling $g_{LQ}=0.1$
(similar effects are found for vector leptoquarks).  For small enough
couplings and large enough masses the curves scale in
$g^2_{LQ}/m_{LQ}^2$. Both constructive and destructive interference
effects, depending on the type of quarks in the final state, are
expected.  The impact of $I=0$, 1 leptoquarks on the hadronic cross
section is larger than the impact of $I=1/2$ leptoquarks.  In the SUSY
interpretation, the HERA events are either $\tilde{c}$ or $\tilde{t}$
production processes with $\lambda'_{121}$ or $\lambda'_{131}\sim
0.05$, respectively. Their impact on \eeqq is shown in Fig.1b for the
total hadronic cross section (solid line). The impact on the charm
quark production $e^+e^-\rightarrow c\bar{c}$ is larger (dotted line),
while it is smaller for the down quarks, $e^+e^- \rightarrow d\bar{d}$ 
(dashed line).

Single leptoquark (with $F=2$) production out of the sea in positron
scattering at HERA requires larger couplings, of order $e$.  For such
couplings leptoquarks with masses $\sim 200$ GeV are excluded by earlier $e^-p $ data and low-energy
limits. In SUSY one can assume $\lambda'_{132}\ne 0$ which
leads to $e^+s\rightarrow \tilde{t}$, or $\lambda'_{123}\ne 0$
which gives rise to  $e^+\bar{c}\rightarrow \tilde{b}^*$ and/or
$e^+b\rightarrow \tilde{t}$ processes. The impact of the former case
on strange quark production at LEP2 is shown in Fig.1b.

Note, that in all SUSY cases we observe only destructive interference
pattern.  Also in contrast to genuine leptoquarks, squarks can decay
via a large number of $R$-parity conserving modes: $\ti{q}\ra q\chi$
with $\chi$ being either a neutralino or a chargino.  If these decays
are non-negligible, the couplings $\lambda'$ would be correspondingly
larger, implying a larger impact on \ee processes.

For large masses, the exchange of leptoquarks/squarks can be described
by contact interactions. Depending on the $LQ/\tilde{q}$ type,
different helicity combinations of lepton and quark currents are
affected differently in either \uu or \dd final states. Potentially
large effects can be expected for \ee annihilation to hadrons for the
scales $\Lambda\sim 2 $ TeV. Present analyses of hadron production at
LEP2 set limits to $\Lambda$ already at the level of about 1.5 to 2.5
TeV~\cite{opal}.

In addition to the lepton-quark-quark superfield term
$\lambda'_{ijk}L^i_LQ^j_L\bar{D}^k_R $, the $R$-breaking part of the 0  0 
superpotential may involve also the interaction of three lepton
superfields $\lambda_{ijk}L^i_LL^j_L\bar{E}^k_R$. Both couplings
$\lambda$ and $\lambda'$ violate lepton number. 
The interpretation of the HERA events by $R$-parity breaking SUSY
interactions involves at least one of the couplings $\lambda'$.  This
invites to specu\-lations that some of the couplings $\lambda$ may also
be non-zero and that other supersymmetric particles, sleptons, may
exist in a similar mass range.  They would influence purely leptonic
processes at LEP2.  Most exciting of course would be the direct
formation of sneutrinos $e^+e^-\ra \ti{\nu}$ and its impact on Bhabha
scattering. This is illustrated in Fig.2, for both (a) virtual 0 
$s$-channel $\tilde{\nu}_{\tau}$ exchange for $m_{\tilde{\nu}_{\tau}}$
exceeding the LEP2 energy, and (b) resonance formation.  The effect
should be also seen in other decay modes of $\tilde{\nu}_{\tau}$
\cite{tau}.

In summary: While the effect of $F=0$ \lqs on \eeqq process at LEP2 is
small, the $F=2$ leptoquarks may lead to observable effects, or at
least more stringent bounds on the Yukawa couplings or contact terms
can be established.  If sleptons do exist in the mass range of 200
GeV, the effect of sneutrino exchanges at LEP2 could be very large.
Most exciting is the prospect that sneutrinos would manifest
themselves through resonance formation in $e^+e^-$ collisions.

\vspace{1.5cm} \hspace*{2mm}
\begin{figure}[htb] 
  \unitlength 0.43mm
\begin{picture}(10,100)(0,0)
  \put(15,19){ \epsfxsize=13.89cm \epsfysize=14.62cm \epsfbox{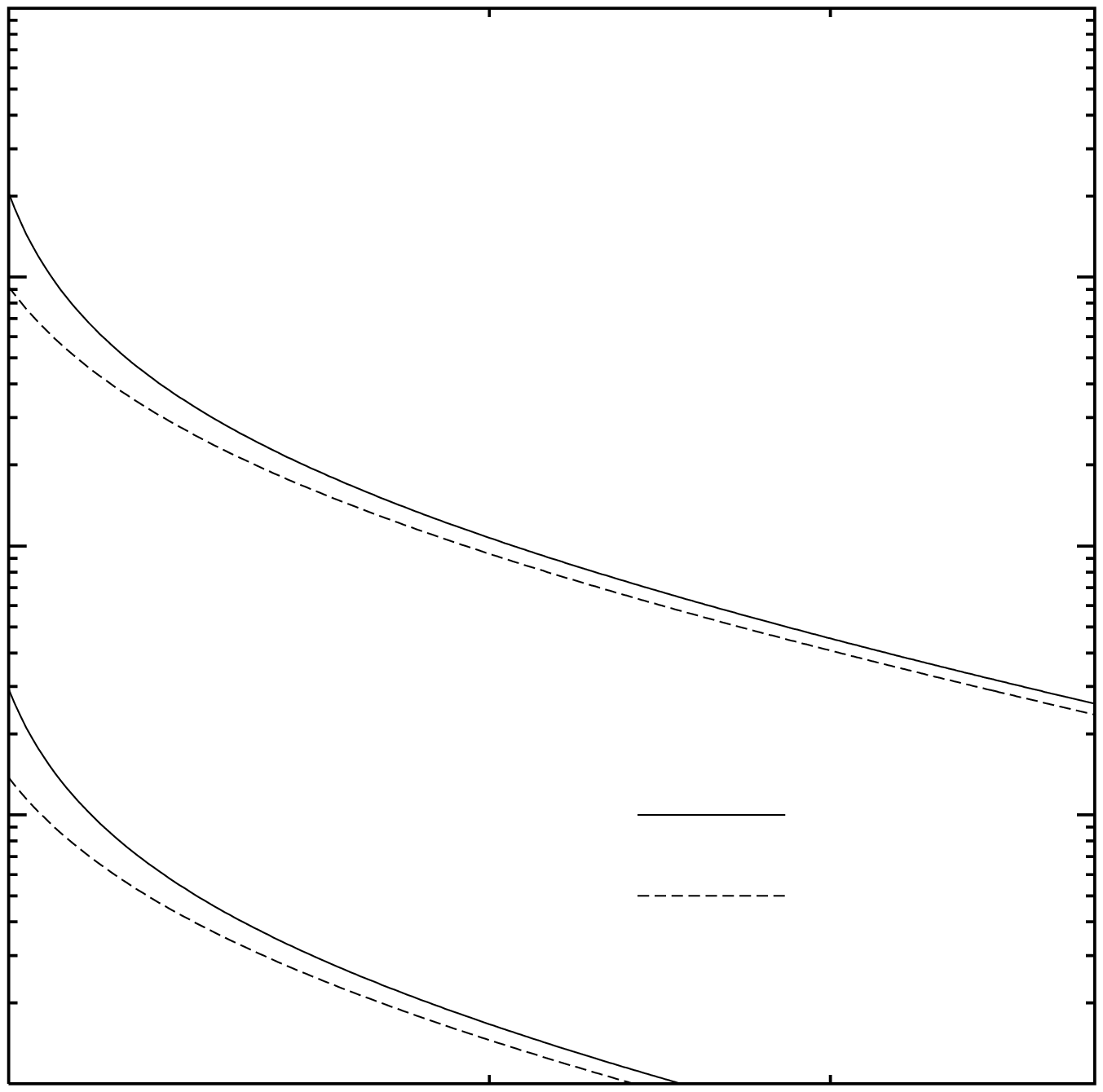}}
  \put(114.6,34){\makebox(0,0)[l]{\small $\protect\sqrt{s} = 184$ }} 
\put(114.6,44){\makebox(0,0)[l]{\small $\protect\sqrt{s} = 192$ }} 
\put(90,92){\makebox(0,0)[l]{\small $45^{\circ} \leq \theta \leq
    135^{\circ}$}} 
\put(30.7,44){\makebox(0,0)[l]{\small $\lambda_{131} = 0.01$}} 
\put(30.7,72){\makebox(0,0)[l]{\small $\lambda_{131} = 0.08$}} 
\put(30,125){\makebox(0,0)[l]{\small $\sigma_{\rm tot}(SM\oplus
    \tilde{\nu}_{\tau})/\sigma_{\rm  tot}(SM)-1$}} 
\put(30,135){\makebox(0,0)[l]{\small $e^+e^- \rightarrow e^+e^-$}} 
\put(126,18){\makebox(0,0){\small $m_{\tilde{\nu}_{\tau}}~$[GeV]}} 
\put(14.7,145){\makebox(0,0)[r]{\small 1}}
\put(14.7,111.5){\makebox(0,0)[r]{\small $10^{-1}$}}
\put(14.7,77.5){\makebox(0,0)[r]{\small $10^{-2}$}}
\put(14.7,44){\makebox(0,0)[r]{\small $10^{-3}$}}
\put(14.7,10.8){\makebox(0,0)[r]{\small $10^{-4}$}}
\put(12.2,5){\makebox(0,0)[l]{\small 200}}
\put(73.7,5){\makebox(0,0)[l]{\small 300}}
\put(116.7,5){\makebox(0,0)[l]{\small 400}}
\put(150.5,5){\makebox(0,0)[l]{\small 500}}
\put(20,20){\makebox(0,0)[l]{(a)}}
\end{picture}
\begin{picture}(100,100)(-150,0)
  \put(15,19){ \epsfxsize=13.89cm \epsfysize=14.62cm
\epsfbox{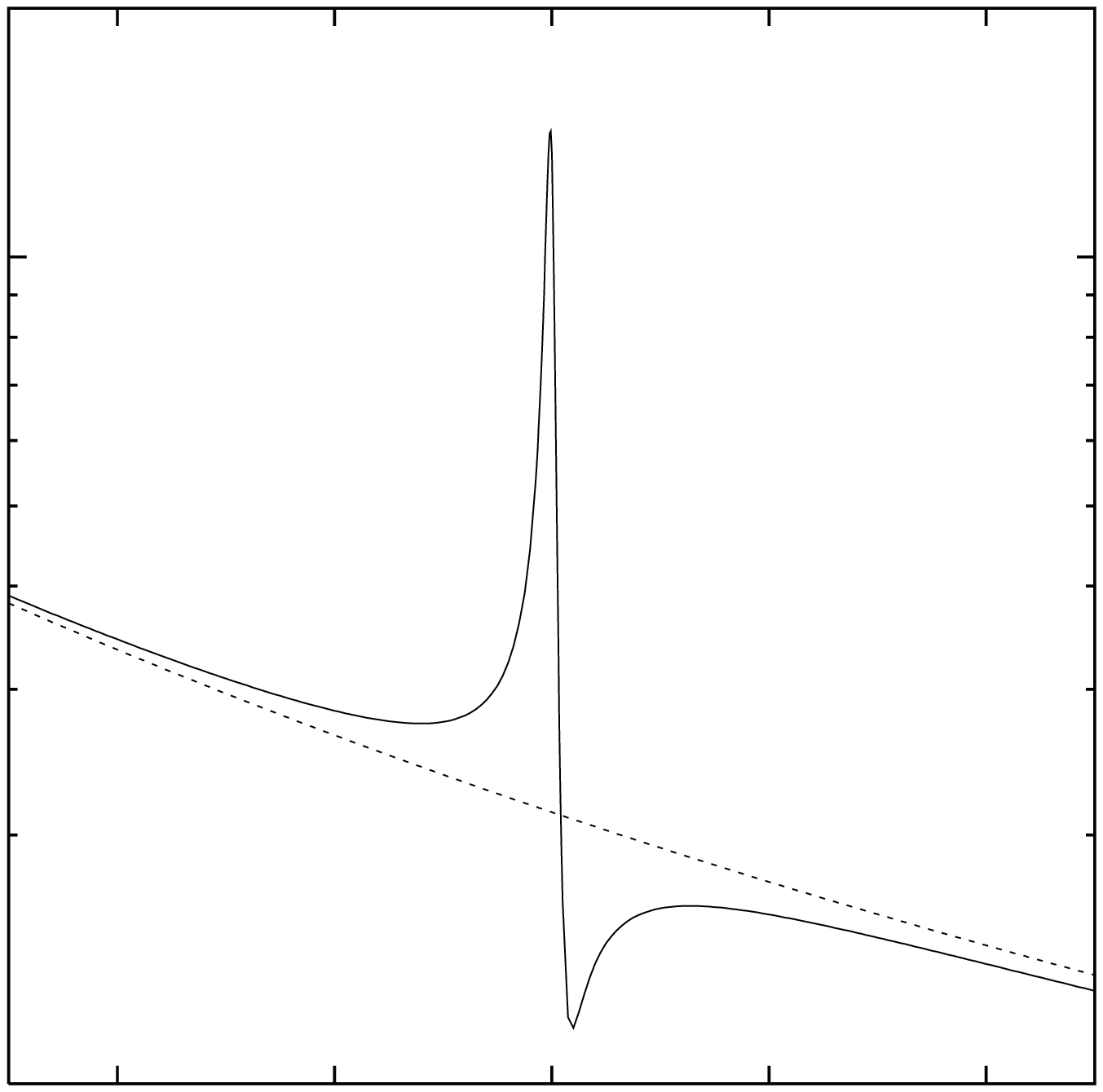}}
\put(25,100){\makebox(0,0)[l]{\small $m_{\tilde{\nu}} = 200$ GeV}}
\put(25,90){\makebox(0,0)[l]{\small $\Gamma_{\tilde{\nu}} = 1$ GeV}}
\put(25,80){\makebox(0,0)[l]{\small $\lambda_{131} = 0.08$}}
\put(95,92){\makebox(0,0)[l]{\small $45^{\circ} \leq \theta \leq
    135^{\circ}$}}
\put(30,135){\makebox(0,0)[l]{\small $\sigma_{\rm tot}(e^+e^-
    \rightarrow e^+e^-)$ [pb]}} 
\put(122,18){\makebox(0,0){\small $\protect\sqrt{s}$ [GeV]}} 
\put(139.3,5.3){\makebox(0,0){\small 240}}
\put(112.0,5.3){\makebox(0,0){\small 220}} 
\put(84.6,5.3){\makebox(0,0){\small 200}}
\put(57.1,5.3){\makebox(0,0){\small 180}} 
\put(29.5,5.3){\makebox(0,0){\small 160}}
\put(14.0,115){\makebox(0,0)[r]{\small $10^2$}}
\put(14.0,10.0){\makebox(0,0)[r]{\small $10$}}
\put(20,20){\makebox(0,0)[l]{(b)}}
\end{picture}
\caption{
  Effect of sneutrino $\ti{\nu}_{\tau}$ exchange on the cross section
  for Bhabha scattering [adapted from J. Kalinowski et al., DESY 97-044, 
hep-ph/9703436].  }
\label{figbb}
\end{figure}

\end{document}